*Original Article*

# Effective Data Aggregation in WSN for Enhanced Security and Data Privacy

B. Murugeshwari[1], S. Aminta Sabatini[2], Lovelit Jose[3], S. Padmapriya[4]

[1,2,3]*Velammal Engineering College, Chennai, India*
[4]*Department of Computer Science & Engineering, Prathyusha Engineering College, Chennai, India*

[1]*Corresponding Author : niyansree@gmail.com*

***Abstract*** *- The two biggest problems with wireless sensor networks are security and energy usage. In sensing devices, malicious nodes could be found in large numbers. The researchers have proposed several methods to find these rogue nodes. To prevent assaults on these networks and data transmission, the data must be secured. Data aggregation aids in reducing the number of messages transmitted within the network, which in turn lowers total network energy consumption. Additionally, when decrypting the aggregated data, the base station can distinguish between encrypted and consolidated analysis based on top of the cryptographic keys. By examining the effectiveness of the data aggregation in this research. To solve the above problem, the system provides a method in which an efficient cluster agent is preferred pedestal on its location at the access point and energy availability. The sensor network's energy consumption is reduced by selecting an effective cluster agent, extending the network's lifespan. The cluster's agent is in indict of compiling data for each member node. The clustering agent validates the data and tosses any errors before aggregation. The clustering agent only aggregates confirmed data. To provide end-to-end anonymity, ElGamal elliptic curve (ECE) encryption is used to secure the client data and reassign the encrypted information en route for the cluster agent. Only the base station (BS) can decrypt the data. Furthermore, an ID-based signature system is utilized to enable authenticity. This research presents a technique for recuperating lost data. The access point employs a cache-based backup system to search for lost data. In the end, the results are contrasted based on many factors, such as encryption and decryption time, aggregation time, and the sum of processing time by adjusting the sensor nodes and altering the cluster agents. The suggested cluster analysis, routing, and security protocol, predicated on the ECE algorithm, significantly outperforms the current practice.*

***Keywords*** *- Wireless Sensor Networks (WSN), Clustering, Routing, Cluster Agent (CA), Security, Data Aggregation.*

## 1. Introduction

In the majority of applications, wireless sensors are used. Memory, power, bandwidth, and inconsistent transmission are wireless sensors' main topics. Wireless sensor network (WSN) processing and storing data is labor-intensive. Therefore, all the data that must be processed separately are maintained in the sensor network, and the rest data is stored in the WSNs that may be utilized in the future. Furthermore, data storage requires a large amount of memory, which can only be distributed over a dispersed wireless network [1].

Additionally, data confidentiality, integrity, and authenticity problems must be addressed when storing large amounts of the data environment. The three features of security offered by wireless sensor networks are discussed here as we talk about storing data in a distributed context. Finally, most of the energy a sensor uses in a wireless network goes toward data processing rather than data storage.

A wireless sensor network (WSN) comprises several nodes powered by batteries, information preparation, and radio communication components. Dynamic wireless sensor networks (WSNs), as opposed to static WSNs, support a more significant focus in terms and more precise administration by enabling the portability of sensor nodes [2]. As a result, dynamic WSNs are typically adopted swiftly in monitoring application processes, mobility streams, automobile status checks, and verifying the health of dairy cattle. Privacy in WNS is one of the most crucial challenges in many fundamental dynamic WSN applications [3-5]. This way, dynamic WSNs must satisfy essential security requirements, including hub authenticity, data-efficient key management in dynamic wireless sensor network anonymity and integrity wherever the nodes move.

Dynamic WSN encryption key management standards based on symmetric encryption have previously been proposed to address security issues. Due to their limited energy and handling capabilities, sensor nodes are a good candidate for this type of encryption. It encounters high connection coverage, though, and needs a lot of memory to hold shared pair-wise keys. It is also unsuitable for aiding hub mobility and is not flexible or versatile versus deals. Symmetric key cryptography is not suitable for dynamic WSNs in this way.

One technique requiring sensors' vitality usage is called data aggregation [6-10]. In such a method, data separated into several parts is summed into a single one by using specific aggregating restrictions, for instance, Average, Sum, and Peak, before being finally communicated to the base station through methods for the wireless connection. As a result, data aggregation helps to decrease circulation and wealth. For example, sensors are transmitted onto the access point in an old forest to provide their felt temperatures for viewing flames [13-15]. In this case, the base station might need to send notifications based on its best estimate of all the identifying data. Each cluster agent needs to choose the most crazily vital capacity from various data values obtained from its component concentrates and then convey the result to the access point shortly after.

There are two or three problems with the sensor design, such as difficulty replacing or recharging the middle batteries due to heavy workloads under pressure and the enigmatic nature of WSNs [16]. One may argue that the most pressing need right now is extending the lifespan of sensing devices, even though it becomes essential, such as lengthening the system lifetime by reducing the criticality of focal point usage in WSNs [39]. The test results show that data exchange is particularly well done based on actual consumption (EC), but contrary to what could be anticipated, side data preparation uses low centrality. A level-agented method was also expected, which would amplify the duration of WSNs and limit the use of large sensors when exchanging data. Another problem with data protection is when data is sent from source to target in WSN.

To extend the lifetime of a sensor network. The data collection for this paper has been enhanced while maintaining a high level of security. The information of the node is continuously updated to the nearby nodes when data is being transferred or forwarded. Euclid distance can be calculated between nodes to identify the neighboring node for packet data forwarding [18]. The lifespan of the infrastructure is extended since less energy is used.

## 2. Related Work

The security mechanism for relaying has indeed been built to guard against man-in-the-middle attacks and access to information vulnerabilities. The current system uses a cryptographic algorithm and a two-column key administration authentication scheme. In this part, many research projects for data aggregation and extending the network lifetime of sensor nodes were covered.

Jenice et al. [5] proposed a Multi-Cluster Secure Information Aggregating Technique used in this research to resolve the two main issues of energy usage and security. First, data between sensor nodes in the WSN is safely aggregating via secure information gathering of various clusters. Data from several sets are encoded using MAC before being collected by an accumulator as part of MCSDA's additive integration of cryptography and multi-data processing. The MCSDA algorithm also uses the fitness value. The method's effectiveness is evaluated in various contexts, and the empirical outcomes are contrasted with specific well-known clustering-based systems. The simulation's findings show that the suggested method performs best in diverse network conditions.

Shraddha et al. [6] suggest the Certificateless-Effective Key Management (CL-EKM) convention. Ensures secure communication in specialized WSNs, illustrated by hub flexibility. The CL-EKM provides forward and reverse essential mystery and supports qualified order to indicate when a center departs or rejoins a network. Moreover, the standard supports practical key renunciation for switched-off nodes and reduces the impact of a hub switch-off on the trustworthiness of other communications joins. Our system has passed a security inspection, which shows that it successfully protects against various attacks. It performs well in terms of time, energy, correspondence, and memory. We also implement a data collection hub for communication within-cluster agents and base stations. When the cluster agent is negotiated hub and information transmission without separation, the DCN is essential.

Dou et al. [7] proposed a classic grouping privacy data-gathering technique. We offer a safe and effective privacy-preserving cluster formation algorithm. In this approach, we embrace the concept of cutting for personal data slicing, construct misleading info for disturbance, and use the SEP protocol to pick ensemble agent nodes proactively. A thorough empirical assessment is carried out to evaluate the effectiveness of the data flow and confidentiality preservation. The outcomes show that the suggested SECPDA algorithm may reduce data traffic and enhance node data privacy.

They offer an elliptic curve-based asymmetric key encryption system, Xiaohan et al. [8] address the problem of WSNs:

- The system improves the key scheme, which periodically produces and handles keys.
- It supports end-to-end cryptography by using privacy homomorphism.
- In order to complete the hop-by-hop authentication, they build a rotating MAC-generating technique.

The experiment results demonstrate that the proposed system uses less energy and provides better security. To improve the system-wide carrying out of tasks by WSNs, Qiyue et al. [9] created an Unmanned Arial Vehicle (UAV) protocol. This method employs an autonomous Ariel device as a data mule to gather sensor data. There are 3 stages to the protocol. The entire network is built initially. The sink then used GA to choose the CHs for each cluster and estimate the pathway for the info mule. The system then moved into the stable phase, during which the data mule traveled the predetermined course and collected data from each cluster. In order to minimize the use of available resources while maintaining data confidentiality and authenticity, Shiva et al. [10] proposed a hybrid algorithm. The hybrid model is used for both encoding and decoding. ECC and AES are the two techniques employed by hybrid algorithms. ECC algorithm is used for cryptography and sharing. Both data encryption processes employ the AES method.

An OSDAP was created by Anish et al. [11] for WSN energy conservation. The encryption is done only at the intermediate node to use the confidentiality additive homomorphic approach. The tree structure divides the data they feel into parts, which are subsequently sent to the current node. The intermediate nodes receive the cipher text from their associated leaf node and combine it with their sensors before decrypting the data. The sink node has to analyze the supplied aggregated data, produce the necessary output for the intended application, and check the accuracy of the data. A Fujisaki Okamoto algorithm created by Nirmal et al. [12] makes the Sybil assault strongly authenticated. It is established as a network with node groups and base stations. In the network, every node has a unique physical ID. The Ad-hoc On-Demand Shortest Path Protocol serves as the route. The base station sends "hello" packets to every other node for architecture verification. The enrolled nodes designate the entire network at the access point.

Anita et al. [13] hypothesized the EDAT technique, the Clusters have a CH depending on their power in the network, and the Q-LEACH protocol selects CH. The whole network categorizes into different groups called "clusters." It uses a MAC-based symmetric encryption algorithm between CH and BS for more secure communication. Finally, a WSN design based on the EDAT algorithm is created to establish encrypted messages between CH and BS. Thus, in the proposed study, EDAT, a highly secure, efficient DA methodology for clarification in WSN, is created, evaluated against other approaches, and hardware-tested. Additionally, this strategy can be expanded by incorporating various attributes while considering more indicators, such as energy recovery techniques.

A study project was suggested by Gulzar et al. [14] to improve the accuracy of remote health monitoring. A design has been designed to generate trust values, and based on the projected value of the trust, a system called CCS with a cryptographic-based approach has been created. This approach aims to improve the accuracy and dependability of the support provided to patients. Furthermore, the intended CCS technique increases precision and trustworthiness using the least energy possible. The mean transmission delay is also kept to a minimum. Finally, a scoring system based on fuzzy logic is employed to show that the suggested scheme outperforms the other approaches.

## 3. Existing System

As a first step, create a network graph in which the nodes carry out the clustering process, divide them into several nodes, and choose a cluster agent randomly for every local network. After that, each node's data center generates routes and distributes keys[19-22]. At every node, the base station's public key is used to encrypt the data. The protected data's hash amount is calculated, and the timestamp is [40]. It sends unique data from each member node to the network agent across all clusters. Next, gather all the info at the cluster agent, check it against the hash value, and either accept it if it is validated or discard it if it is not. All data is aggregated during the aggregating phase, and this information is sent to the access point [24-28]. When the central server receives data from each cluster agent, it checks the data and uses the proper key to decode the data

## 4. Proposed Methodology

A network is originally founded with vertices and nodes densely interconnected. Following network formation, a method known as clustering is carried out in which nodes are split up into various clusters. The clustering agent is chosen from each group of interconnected clusters after the collections have been generated based on energy, proximity to the access point, and neighboring nodes. Each node gets a key from the base station, which also transmits keys to the other nodes. Generating routes from each node to the access point is to be done.

The elliptic curve ElGamal (ECE) cryptosystem produces and encrypts data for each node. Once the data has been encoded, the hash function is evaluated, and the timestamp is logged. Each node transmits data to its cluster agent after analyzing the hash value on-site. Cluster agents gather all the data and ensure that it is accurate. The clustering agent confirms the validity of the data before concluding the data aggregation procedure. Data to the access point, too. Every cluster agent transmits data to a central, decrypting it using the proper key [41]. The base station searches for missing data and uses a perfecting backup system to find it.

### 4.1. Process Description
The suggested methodology represents a system's flow. First, an infrastructure of sensor nodes is established. Next, a clustering algorithm is applied, and the number of sensor nodes is partitioned into clusters. A cluster agent is then chosen based on three criteria. Key distribution is then carried out with each node via the base station, and a route is then produced from every node to the BS. Secure the data using the secret key and the elliptic curve ElGamal (ECE) algorithm. The encrypted data's hash value is determined, and a timestamp is noted. Each cluster's cluster agent receives data from the cluster member. Data is validated using its hash value; if validated, it is accepted; else, it is discarded. Then combine every data from the node and send it to the BS. Using the proper keys, the central server decodes the data [30]. The base station runs the cache-based retrieval method for data restoration and examines it for missing data, as displayed in Figure 1.

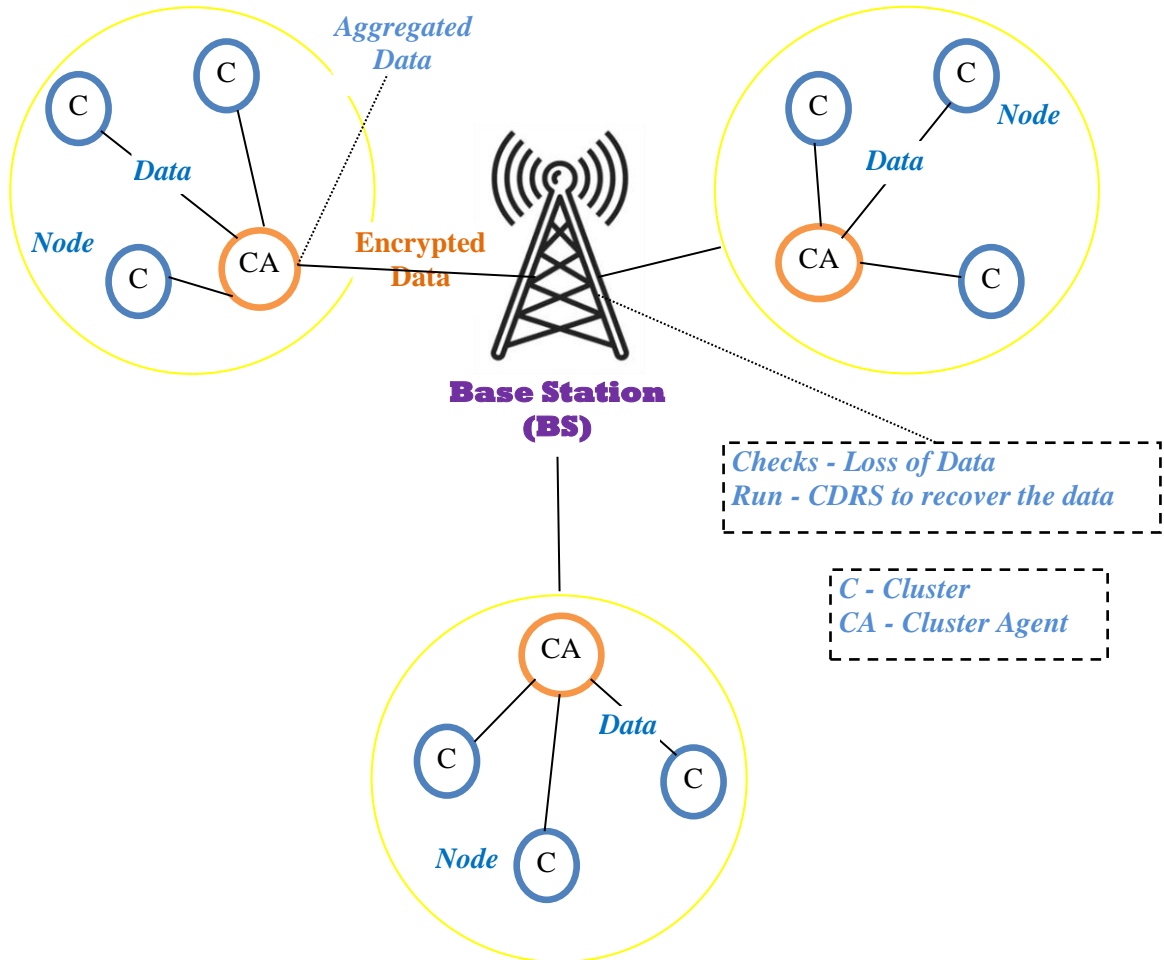

**Fig. 1 Architecture of proposed model**

### 4.2. Initialization of Nodes

1. Initially, randomly choose s1 -- > $Z^*_q$

2: $PK = s_1P$, $s_2 = x \cdot H1(id)$

3: Then, randomly selects key - k from $Z^*_q$

4: Calculate $r = kP + PK$

5: Convert --> $CT_1 = h(PK, r)$ and $CT_2 = k - x \cdot CT_1$

6: pre-stack the data node with ($s_1$, $s_2$, $PK$, $CT_1$, $CT_2$, param) and the energy

7: a= 0

8: Make NKList =empty

9: Make Valid =true

### 4.3. Proposed Algorithm STEPS

Step 1. Graph g(v,e) is used to create a network graph, where V stands for vertices and E for edges.

Step 2. Use the classification approach to separate the nodes into a specific count of clusters.

Step 3. Select the Effective Cluster Agent based on the energy, the set of nodes, and the distances to the access point.

Step 4. Distribute the keys at each node using the base station.

Step 5. Generating routes from every node to the access point is to be done.

Step 6. Create data at each node, then secure it using the base station's decryption key.

Step 7. Determine the encoded data's hash value and note the timestamp.

Step 8. Out of each cluster member across all clusters, provide the personal information to the cluster agent.

Step 9. At the cluster agent, gather all the data. Validate the information using its hash value, then either approve it after verification or reject it if the hash function is incorrect.

Step 10. Assemble all the information, then send it to the access point. Each cluster agent submits data to the BS.

Step 11. BS checks the information and uses the proper key to decode it.

Step 12. The base station searches for lost data and uses a cache-dependent recovery method to restore lost data.

### 4.4. EC ElGamal Cryptography

Inelastic curve An asymmetric cryptographic technique is ElGamal(EC). Using a crypto encryption key has a known benefit [31-35]. With the help of the crypto elliptic curve, the message is mapped (CE). Transform plaintext with CE first before encrypting a text using EC. The plaintext t is compounded by point P to get the CE point Tp using a straightforward mapping process. The plaintext is equivalent to CE in terms of addition.

**Encryption Process**

**Input:** key - Public PK, plaintext pt

**Output:** cipher text (C, CT)

step 1: choose the random key - $Pk \epsilon [1, pt-1]$

Step 2: M = map(pt)

Step 3: C = fPt

Step 4; CT = Pt + kPK

Step 5: return (C, CT)

**Decryption Process**

**Input:** Private key PrK, cipher text(C, CT)

**Output:** plaintext pt

Step 6: M = PrK.C + CT

Step 7: CT = map(M)

Step 8: return the pt

## 5. Clustering and Cluster Agent Selection

The "ai" nodes in this clustering-based approach located in the same region share the exact cluster ci and a special attribute Atr. Cluster agent CAi, one of the cluster members, is chosen to gather and encode the data with cluster attribute Atr and transmit it to the BS via the closest RCA with the identification number IDnr. As a result, we choose a CA in each round based on the following three parameters to manage the network lifetime.

### 5.1. Maximum Residual Energy

To increase the system's lifespan, maintain consistent data transmission to the BS, and balance the load on power indulgence, choose the CA with the highest remaining energy among some cluster members. Depending on the research length and the distance, nodes use a certain amount of energy when sending and receiving data. The free space and the multi-path rayleigh fading models with boosting index are employed, respectively, and EDA indicates the energy consumed by data fusion. The radio dispersion energy consumed by transmitting data and retrieving files is stated as Em.

A node uses the following amount of energy when sending a 1-bit payload across a length of "ai":

$$E_{\text{Trans}}(len, v) = E_{Trans-elec}(len) + E_{\text{Trans}-amp}(len, v) = E_m * len + \epsilon_{fs} v^2 * len \quad v < v0$$
$$E_m * len + \epsilon_{fs} + \epsilon_{amp} * v^4 * len \quad v \geq v0$$

The amount of energy used to receive this message is:

$$E_{Rec}(l) = E_m * len$$

### 5.2. Minimal Distance between Nodes and Bases

The distance between CA and BS should be as small as possible to reduce energy consumption. In order to expand reachability to BS, providing BS at the second level adds one hop.

Distance between node and BS = $\sum_{i=1}^{n}(Xi - Yi)^2$

The parameters on the x and y axes for each node to BS are Xi and Yi.

### 5.3. Maximum Number of Nodes in the Transmission Range of that Node

For the cluster agent to gather data from every cluster participant, as many nodes as feasible must be closest to it. All member nodes transfer data to the relay cluster agent if the cluster agent is disrupted.

#### 5.3.1. Combining Secure Data

While addressing secure data aggregation, WSN also covers the collection results by the network's aggregate networks or supervisory nodes. Key security considerations include data integrity and confidentiality. Various issues with the current algorithms cause heavy computation owing to inefficiencies and connections over agents when aggregating secure data in WSN [42]. When security flaws are found, this issue arises. Furthermore, the current algorithms do not combine data aggregation, reliability, privacy, and fake data recognition. The suggested method quickly identifies the bogus data. With this method, transmission coverage can be reduced, low energy usage is achieved, and the lifespan of significant network systems is increased. To face the security problem in the WSN - wireless sensor network, this article presents a new secure data aggregation approach based on cryptographic encryption. The ElGamal data encryption algorithm is paired with MAC to ensure data integrity in the aggregate of a specific data strategy, which safeguards the integrity and confidentiality of end-to-end data, as displayed in Figures 2 and 3. The following steps comprise a secure distributed processing model: key creation, encryption, MAC technique, aggregate, and authentication. The application of these procedures is divided into three areas.

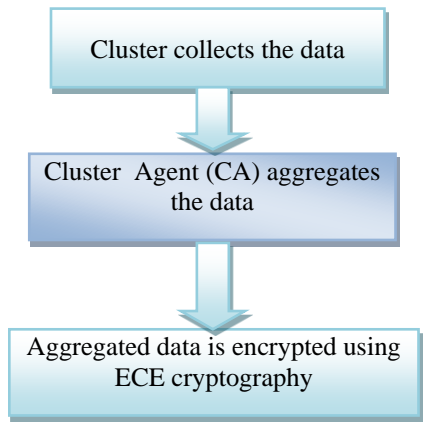

**Fig. 2 Flow diagram for Transmitter section in CA**

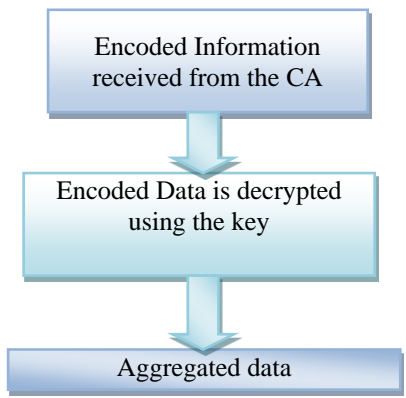

**Fig. 3 Flow diagram for Receiver section in the Internet (BS)**

MAC is used in the clustering process to safeguard the privacy and accuracy of all data. A message authentication code is a crypto mechanism that provides authentication services. The little piece of data needed to verify that the intended recipient sent the message and that it was not manipulated. MAC guarantees a message's legitimacy and integrity are guaranteed by MAC, which enables timestamping to see any modifications to the message's content. The MAC method validates a transmission and principally tries to split data packets into manageable chunks. When a message or piece of data is sent to a recipient with an authentication server (a set of key values), it is said to be authenticated. e transmitter and recipient must share a key, called k, to create the MAC process. Key k chooses a key at random from the space during key generation. The nodes enter the data and the secret keys ks into the MAC algorithm to produce a MAC.

The effective signing procedure produces a MAC from the content and the key. The nodes forward the data together with the MAC. The authenticity and anonymity of the message's origin are crucial issues in the message sent to the collector. If data privacy is necessary, the data must be encrypted. The aggregator updates the MAC value after receiving the message and MAC by sending the message obtained and the key that was transferred to the MAC algorithm [43]. During the verification phase, the aggregator successfully confirms the message's authentication by determining whether or not it is duplicated. The aggregator now checks to see if the received MAC and the obtained values MAC from the transmitter are equal.

The message will be acknowledged by the aggregator and forwarded by the target node if the obtained message is satisfied. The aggregator determines whether the message is updated or has a fake source if the MAC sent by the nodes does not match the calculated MAC. Finally, the receiver comes to the firm conclusion that the message is fake. The Aggregation site sends a message to the base station if it is authentic. Message digests, also known as MACs, are used to safeguard the integrity of a piece of data or media by detecting modifications and alterations to any component of a message. Ks serve as a standard key for encryption. Preserving a text's legitimacy, confirming an originator's identity, and maintaining the origin are all goals of text authenticity.

## 6. Performance Analysis

The effectiveness assessment is carried out to see how well the suggested procedure performs compared to other protocols. NS-2 is the simulation platform that is most frequently utilized. The suggested protocol's performance analysis is simulated using a network simulator. This section begins by providing a succinct definition of measuring performance. Following that, this part provides a brief explanation of the simulation platform and the different testing parameters. The comparison will be demonstrated after thoroughly analyzing the proposed protocol's result. The effectiveness of the results is comparable to that of the CH Trust Propagation Scheme (CHTPS) and the Efficient and Provably Secure Aggregation (EPSA) protocols.

## 6.1. Performance Metrics

According to the current research, the strategy is carried out on a computer system with an Intel Core i5 CPU running at 2.66 GHz and 4 GB of RAM. Although the CPU performs quicker than actual sensor nodes, it can still be utilized to show how well-allied approaches perform.

The encrypt, decrypt, aggregate, and overall calculation time of the proposed and conventional techniques are displayed in Figures 4, 5, and 6 accordingly. The suggested methodology is compared with the previous algorithm using several estimating criteria, such as the Efficient Data Aggregation Technique (EDAT) design and the Efficient and Provably Secure Aggregation (EPSA) method.

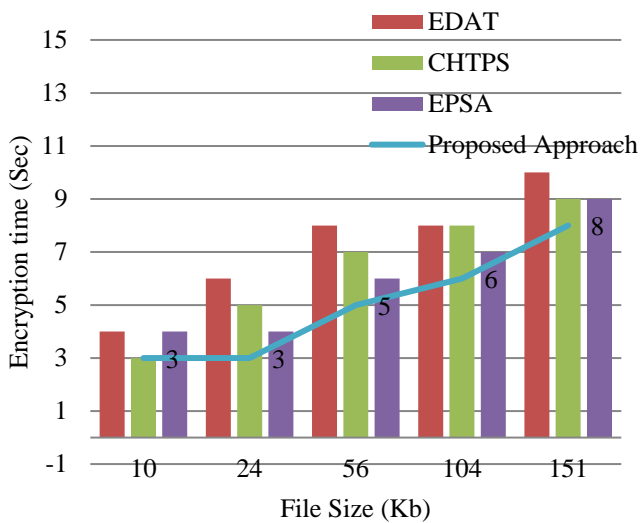

**Fig. 4 Encryption time**

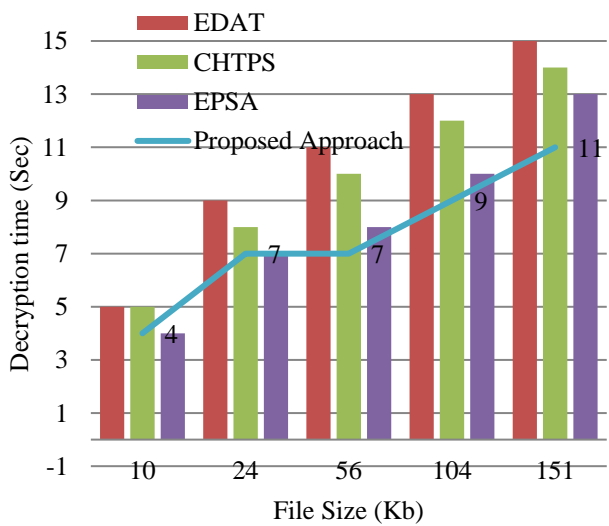

**Fig. 5 Decryption time**

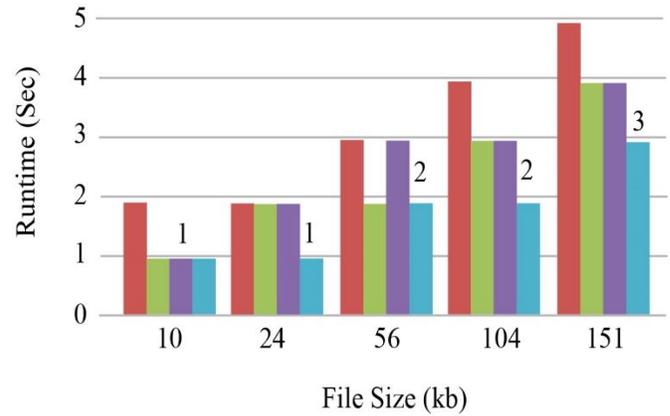

**Fig. 6 Aggregation time**

In contrast, Figure 7 depicts the whole calculation time for a single cluster agent, changing sensor nodes. Figure 8 displays the total computational effort for 20 sensor nodes, 10 kB of file size, and changing the cluster Agents. As SIES lacks an aggregate signature, it requires additional calculation time in both scenarios. Because EPSA includes more header information at each level, it takes longer to compute in both scenarios. The energy used for varying the sensor nodes and the calculation time is directly correlated [38]. If it is lower, the sensor nodes' need for energy will also be lower. A device's energy usage is determined using the formula below.

$$E = V.I / T$$

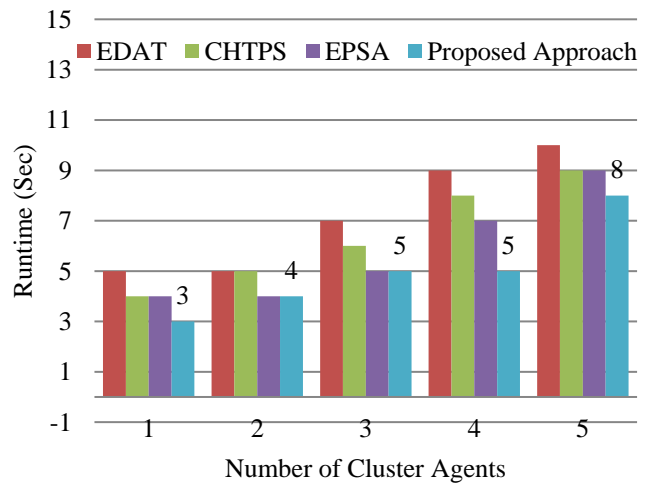

**Fig. 7 Overall processing time when the cluster agents are changed.**

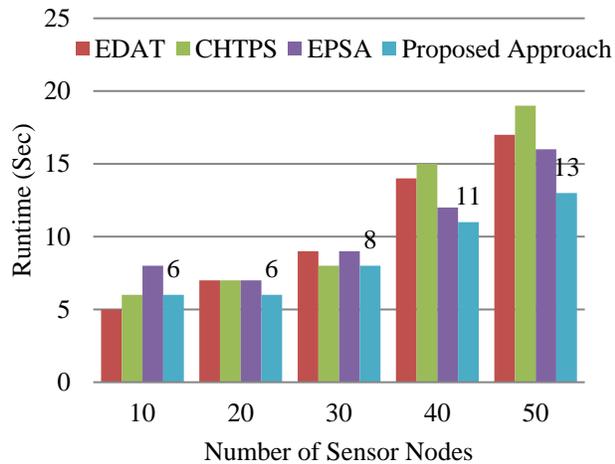

**Fig. 8 Overall processing time when the sensor nodes are changed**

Where E is energy, V is voltage, I is current, and T is time. For every hardware, V and I will always be constant. The CPU's energy consumption is considered when calculating the energy because the implementation is done on a computing device. The current and voltage for the CPU can be considered to be 3 V and 8 mA, correspondingly. As a result, E = 3 0.0 is the computed power consumption for encrypting 10 kB of data. Similar calculations are made for other methods with different data sizes.

## 7. Conclusion

In the current study, secure information aggregating encrypted files is accomplished on WSNs. At the same time, data Security and validity are guaranteed by composite authentication. Data secrecy is given through the use of encryption. Despite the higher expense, aggregate verification is still appropriate for WSNs. Comparing the performance of the proposed system to other similar systems in terms of computing time and energy usage shows that it is practical for protecting information aggregation. To prevent assaults on these networks and data transmission, the data must be secured. Data aggregation aids in reducing the number of messages transmitted within the network, which in turn lowers total network energy consumption, by examining the effectiveness of data aggregation in this research. To solve this problem, the system provides a method in which an efficient cluster agent is preferred depending on their location for the access point and energy availability. The sensor network's energy consumption is reduced by selecting an effective cluster agent, extending the network's lifespan. The cluster's agent is in charge of compiling data from each member node. The clustering agent validates the data and tosses any errors before aggregating. The clustering agent only aggregates confirmed data.

ElGamal elliptic curve (ECE) encryption is used to secure the data and transfer the encrypted information to the cluster agent to provide end-to-end anonymity. The results are contrasted based on many factors, such as Encryption and Decryption Time, Aggregation Time, and the sum of calculation time by changing the sensor and the cluster agents. The suggested cluster analysis, routing, and security protocol, predicated on the ECE algorithm, significantly outperforms the current protocol.

## References

[1] Khalid A. Darabkh, Mohammad Z. El-Yabroudi, and Ali H. El-Mousa, "BPA-CRP: A Balanced Power-Aware Clustering and Routing Protocol for Wireless Sensor Networks," *Ad Hoc Networks,* vol. 82, pp. 155-171, 2019. Crossref, https://doi.org/10.1016/j.adhoc.2018.08.012

[2] B. Murugeshwari, D. Selvaraj, K. Sudharson, and S. Radhika, "Data Mining with Privacy Protection using Precise Elliptical Curve Cryptography," *Intelligent Automation & Soft Computing,* vol. 35, no.1, pp. 839–851, 2023. Crossref, https://doi.org/10.32604/iasc.2023.028548

[3] Taochun Wang, Xiaolin Qin, Youwei Ding, Liang Liu, and Yonglong Luo, "Privacy-Preserving and Energy-Efficient Continuous Data Aggregation Algorithm in Wireless Sensor Networks," *Wireless Personal Communications*, vol. 98, no. 1 pp. 665-684, 2018. Crossref, https://doi.org/10.1007/s11277-017-4889-5

[4] B. Murugeshwari, K. Sarukesi and C. Jayakumar, "An Efficient Method for Knowledge Hiding Through Database Extension," *Test Conference International,* pp. 342-344, 2010. Crossref, https://doi.org/10.1109/ITC.2010.93

[5] A, Jenice and D, Hevin, "An Energy Efficient Secure Data Aggregation in Wireless Sensor Networks," *Research Square,* 2021. Crossref, https://doi.org/10.21203/rs.3.rs-364741/v1

[6] Shraddha Deshmukh, A. R. Bhagat Patil and Harshad Nakade, "Implementation of Effective Key Management Strategy with Secure Data Aggregation in Dynamic Wireless Sensor Network," *International Journal of Scientific Research in Computer Science, Engineering and Information Technology*, vol. 4, no. 2, pp. 358-364, 2018.

[7] Dou H, Chen Y and Yang Y, "A Secure and Efficient Privacy-Preserving Data Aggregation Algorithm," *Journal of Ambient Intelligence and Humanized Computing,* vol. 13, pp. 1495–1503, 2022. Crossref, https://doi.org/10.1007/s12652-020-02801-6

[8] Xiaohan Qi, Xiaowu Liu, Jiguo Yu and Qiang Zhang, "A Privacy Data Aggregation Scheme for Wireless Sensor Networks," *Procedia Computer Science*, vol. 174, pp. 578-583, 2020. Crossref, https://doi.org/10.1016/j.procs.2020.06.127

[9] Qiyue Wu, Peng Sun, and Azzedine Boukerche, "An Energy-Efficient UAV-Based Data Aggregation Protocol in Wireless Sensor Networks," In *Proceedings of the 8th ACM Symposium on Design and Analysis of Intelligent Vehicular Networks and Applications*, pp. 34-40, 2018. Crossref, https://doi.org/10.1145/3272036.3272047

[10] Shiva Prakash and Ashish Rajput, "Hybrid Cryptography for Secure Data Communication in Wireless Sensor Networks," In *Ambient Communications and Computer Systems*, *Springer*, pp. 589-599, 2018. Crossref, https://doi.org/10.1007/978-981-10-7386-1_50

[11] Anish Soni and Dr.Rajneesh Randhawa, "OSDAP-Optimized and Secure Data Aggregation Protocol for Wireless Sensor Networks," *International Journal of Applied Engineering Research,* vol. 13, no. 5, pp. 3027-3033, 2018.

[12] K. Nirmal Raja and M. Maraline Beno, "Secure Data Aggregation in Wireless Sensor Network-Fujisaki Okamoto (FO) Authentication Scheme Against Sybil Attack," *Journal of Medical Systems,* vol. 41, no. 7, pp. 107, 2017. Crossref, https://doi.org/10.1007/s10916-017-0743-2

[13] D. Anita Daniel, S. Emalda Roslin, "An Efficient Data Aggregation Technique for Green Communication in WSN," *International Journal of Engineering Trends and Technology*, vol. 69, no. 3, pp. 138-146, 2021. Crossref, https://doi.org/10.14445/22315381/IJETT-V69I3P222

[14] G. Mehmood, M. Z. Khan, A. Waheed, M. Zareei and E. M. Mohamed, "A Trust-Based Energy-Efficient and Reliable Communication Scheme (Trust-Based ERCS) for Remote Patient Monitoring in Wireless Body Area Networks," in IEEE Access, vol. 8, pp. 131397-131413, 2020. Crossref, https://doi.org/10.1109/ACCESS.2020.3007405

[15] Dhinakaran D and Joe Prathap P. M, "Protection of Data Privacy from Vulnerability Using Two-Fish Technique with Apriori Algorithm in Data Mining," *The Journal of Supercomputing*, vol. 78, pp. 17559–17593, 2022. Crossref, https://doi.org/10.1007/s11227-022-04517-0

[16] K. Sudharson, M. Akshaya, M. Lokeswari and K. Gopika, "Secure Authentication Scheme Using CEEK Technique for Trusted Environment," *2022 International Mobile and Embedded Technology Conference (MECON),* pp. 66-71, 2022. Crossref, https://doi.org/10.1109/MECON53876.2022.9752245

[17] D. Anita Daniel and S. Emalda Roslin, "An Efficient Data Aggregation Technique for Green Communication in WSN," *International Journal of Engineering Trends and Technology,* vol. 69, no. 3, pp. 138-146, 2021. Crossref, https://doi.org/10.14445/22315381/IJETT-V69I3P222

[18] Jeyalakshmi. C, Balasubramaniam, Murugeshwari and Karthick, M, "HMM and K-NN based Automatic Musical Instrument Recognition," *IEEE,* pp. 350-355, 2018. Crossref, https://doi.org/10.1109/I-SMAC.2018.8653725

[19] Khalid A. Darabkh, Saja M. Odetallah, Zouhair Al-qudah, Khalifeh Ala'F, and Mohammad M. Shurman, "Energy-Aware and Density-Based Clustering and Relaying Protocol (EA-DB-CRP) for Gathering Data in Wireless Sensor Networks," *Applied Soft Computing*, vol. 80, pp. 154-166, 2019. Crossref, https://doi.org/10.1016/j.asoc.2019.03.025

[20] Mohamed Elshrkawey, Samiha M. Elsherif, and M. Elsayed Wahed, "An Enhancement Approach for Reducing The Energy Consumption In Wireless Sensor Networks," *Journal of King Saud University-Computer and Information Sciences*, vol. 30, no. 2, pp. 259-267, 2018. Crossref, https://doi.org/10.1016/j.jksuci.2017.04.002

[21] K. Sudharson and S. Arun, "Security Protocol Function Using Quantum Elliptic Curve Cryptography Algorithm," *Intelligent Automation & Soft Computing*, vol. 34, no. 3, pp. 1769–1784, 2022. Crossref, https://doi.org/10.32604/iasc.2022.026483

[22] Balasubramaniam Murugeshwari, Daniel Raphael and Raghavan Singaravelu, "Metamaterial Inspired Structure with Offset-Fed Microstrip Line for Multi-Band Operations," *Progress in Electromagnetics Research*, vol. 82, pp. 95-105, 2019.

[23] Vinod Kumar and Om Prakash Roy, "A Reliable and Secure Inter-and Intra-State Routing Protocol for VoIP communication," *International Journal of Engineering Trends and Technology,* vol. 70, no. 7, pp. 479-490, 2022. Crossref, https://doi.org/10.14445/22315381/IJETT-V70I7P250

[24] Tianshu Wang, Gongxuan Zhang, Xichen Yang, and Ahmadreza Vajdi, "Genetic Algorithm for Energy- Efficient Clustering and Routing in Wireless Sensor Networks," *Journal of Systems and Software,* vol. 146, pp. 196-214, 2018. Crossref, https://doi.org/10.1016/j.jss.2018.09.067

[25] D. Dhinakaran, P.M. Joe Prathap, D. Selvaraj, D. Arul Kumar, and B. Murugeshwari, "Mining Privacy-Preserving Association Rules based on Parallel Processing in Cloud Computing," *International Journal of Engineering Trends and Technology,* vol. 70, no. 3, pp. 284-294, 2022. Crossref, https://doi.org/10.14445/22315381/IJETT-V70I3P232

[26] Avinash Rai, and Preetu Patel, "Secure Data Aggregation Technique for Wireless Sensor Networks using Iterative Filtering Algorithm," *International Journal of Applied Engineering Research,* vol. 13, no. 18, pp. 13805-13811, 2018.

[27] D. Dhinakaran, D. A. Kumar, S. Dinesh, D. Selvaraj and K. Srikanth, "Recommendation System for Research Studies Based on GCR," *2022 International Mobile and Embedded Technology Conference (MECON),* Noida, India, pp. 61-65, 2022. Crossref, https://doi.org/10.1109/MECON53876.2022.9751920

[28] J. Aruna Jasmine, V. Nisha Jenipher, J. S. Richard Jimreeves, K. Ravindran, and D. Dhinakaran, "A Traceability Set Up Using Digitalization of Data and Accessibility," *2020 3rd International Conference on Intelligent Sustainable Systems (ICISS)*, pp. 907-910, 2020. Crossref, https://doi.org/10.1109/ICISS49785.2020.9315938

[29] M.Supriya, Dr.T.Adilakshmi, "Secure Routing using ISMO for Wireless Sensor Networks," *SSRG International Journal of Computer Science and Engineering*, vol. 8, no. 12, pp. 14-20, 2021. Crossref, https://doi.org/10.14445/23488387/IJCSE-V8I12P103

[30] K.Sudharson, Ahmed Mudassar Ali and N.Partheeban, "NUITECH – Natural user Interface Technique Foremulating Computer Hardware," *International Journal of Pharmacy & Technology,* vol. 8, no. 4, pp. 23598-23606, 2016.

[31] S. Arun and K. Sudharson. "DEFECT: discover and eradicate fool around node in Emergency Network using Combinatorial Techniques," *Journal of Ambient Intelligence and Humanized Computing,* pp. 1-12, 2020. Crossref, https://doi.org/10.1007/s12652-020-02606-7

[32] J. A. Shanny and K. Sudharson, "User Preferred Data Enquiry System using Mobile Communications," *International Conference on Information Communication and Embedded Systems (ICICES2014),* pp. 1-5, 2014. Crossref, https://doi.org/10.1109/ICICES.2014.7033943

[33] D. Dhinakaran and P.M. Joe Prathap, "Ensuring Privacy of Data and Mined Results of Data Possessor in Collaborative ARM, Pervasive Computing and Social Networking," *Lecture Notes in Networks and Systems, Springer, Singapore,* vol. 317, pp. 431-444, 2022. Crossref, https://doi.org/10.1007/978-981-16-5640-8_34

[34] D. Dhinakaran and P. M. Joe Prathap, "Preserving Data Confidentiality in Association Rule Mining Using Data Share Allocator Algorithm," *Intelligent Automation & Soft Computing,* vol. 33, no. 3, pp. 1877-1892, 2022. Crossref, https://doi.org/10.32604/iasc.2022.024509

[35] K. Sudharson, A. M. Sermakani, V. Parthipan, D. Dhinakaran, G. Eswari Petchiammal and N. S. Usha, "Hybrid Deep Learning Neural System for Brain Tumor Detection," *2022 2nd International Conference on Intelligent Technologies (CONIT),* pp. 1-6, 2022. Crossref, https://doi.org/10.1109/CONIT55038.2022.9847708

[36] K.Priyadharshini, "A Review on Wireless Sensor Networks-Security Issues and Disputes," *International Journal of P2P Network Trends and Technology,* vol. 9, no. 2, pp. 6-9, 2019.

[37] Korada Kishore Kumar and Konni Srinivasa Rao, "An Efficient users Authentication and Secure Data Transmission of Cluster-based Wireless Sensor Network," *SSRG International Journal of Computer Science and Engineering*, vol. 5, no. 1, pp. 1-5, 2018. Crossref, https://doi.org/10.14445/23488387/IJCSE-V5I1P101

[38] Khalid A. Darabkh, Noor J. Al-Maaitah, Iyad F. Jafar and Ala'F Khalifeh, "EA-CRP: A Novel Energy-Aware Clustering and Routing Protocol in Wireless Sensor Networks," *Computers & Electrical Engineering,* vol. 72, pp. 702-718, 2018. Crossref, https://doi.org/Crossref, https://doi.org/10.1016/j.compeleceng.2017.11.017

[39] Murugeshwari, "Preservation of the Privacy for Multiple Custodian Systems with Rule Sharing," *Journal of Computer Science,* vol. 9, no. 9, pp. 1086-1091, 2013. Crossref, https://doi.org/10.3844/jcssp.2013.1086.1091

[40] B. Murugeshwari, R.S. Daniel and S. Raghavan, "A Compact Dual Band Antenna Based on Metamaterial-Inspired Split Ring Structure and Hexagonal Complementary Split-Ring Resonator for ISM/Wimax/WLAN Applications," *Applied Physics A,* vol. 125, pp. 628, 2019. Crossref, https://doi.org/10.1007/s00339-019-2925-x

[41] N.Partheeban, K.Sudharson and P.J.Sathish Kumar, "SPEC- Serial Property Based Encryption for Cloud," *International Journal of Pharmacy & Technology,* vol. 8, no. 4, pp. 23702-23710, 2016.

[42] K. Sudharson and V. Parthipan, "A Survey on ATTACK – Anti-Terrorism Technique for Adhoc Using Clustering and Knowledge Extraction," *Advances in Computer Science and Information Technology*, Computer Science and Engineering, CCSIT 2012, Lecture Notes of the Institute for Computer Sciences, Social Informatics and Telecommunications Engineering, Springer, Berlin, Heidelberg, vol. 85, pp. 508-514, 2012. Crossref, https://doi.org/10.1007/978-3-642-27308-7_54

[43] K. Sudharson and V. Parthipan, "SOPE: Self-Organized Protocol for Evaluating Trust in MANET Using Eigen Trust Algorithm," *2011 3rd International Conference on Electronics Computer Technology*, pp. 155-159, 2011. Crossref, https://doi.org/10.1109/ICECTECH.2011.5941675